# Marangoni-Driven Redistribution and Activity of Piezo1 Molecules in Epithelial and Cancer Cells

Ivana Pajic-Lijakovic [1,4], Milan Milivojevic [1], Boris Martinac [2,3], and Peter V. E. McClintock [4]

[1] University of Belgrade, Faculty of Technology and Metallurgy, Department of Chemical Engineering, Belgrade, Serbia

[2] Mechanosensory Biophysics Laboratory, Victor Chang Cardiac Research Institute, Sydney, Australia

[3] St Vincent's Clinical School, Faculty of Medicine, University of New South Wales, Sydney, New South Wales, Australia

[4] Department of Physics, Lancaster University, Lancaster LA1 4YB, UK

Correspondence to:    Ivana Pajic-Lijakovic, iva@tmf.bg.ac.rs and,

Peter V. E. McClintock, p.v.e.mcclintock@lancaster.ac.uk

**Abstract**

The activity and distribution of Piezo1 molecules, along with the maturity and strength of focal adhesions (FAs), serve as critical factors influencing cell mechanosensing. Notably, migrating epithelial cells and mesenchymal-like cancer cells exhibit significantly different behaviors regarding these elements. In cancer cells, Piezo1 molecules are distributed uniformly, while in epithelial cells, their distribution is heterogeneous. In epithelial cells, Piezo1 molecules tend to group around FAs, a phenomenon that is enhanced by actomyosin contractility. However, a reduction in contractility results in a more uniform distribution of Piezo1 molecules. The expression and activity levels of Piezo1 molecules are markedly higher in cancer cells compared to epithelial cells. The activity of Piezo1 molecules correlates with the intracellular calcium concentration. Despite the extensive experimental studies on the properties of migrating epithelial and mesenchymal-like cancer cells, the physical explanations remain lacking. The primary objective of this theoretical study is to explore: (i) the inhomogeneous distribution of Piezo1 molecules in epithelial cells in relation to the Marangoni effect, (ii) the heightened activity of Piezo1 molecules in cancer cells by specifying the driving force, and (iii) the influence of membrane-mediated interactions among Piezo1 molecules grouped near FAs in epithelial cells on their activity.

**Glossary of terms**

**Adherens junction:** is a type of cell-cell adhesion contact, which plays an important role for migrating cell collectives. Cells use AJs for mechanical coupling and as an important source of signalling that coordinates collective behaviour.

**Anisotropy:** is the property of a system where its physical characteristics vary depending on the direction in which they are measured. This means a system can exhibit different properties, like strength or stiffness, along different axes.

**Cadherins**: are transmembrane glycoproteins containing an extracellular domain that mediates cell-cell adhesion via homophilic or heterophilic interaction, and an intracellular domain that controls signaling cascades involved in a variety of cellular processes, including polarity, gene expression, etc.

**Cell mechanosensing:** is the ability of cells to detect and respond to mechanical cues—such as stiffness, tension, compression, or shear—through specialized molecular structures like integrins, the cytoskeleton, and mechanosensitive ion channels.

**Epithelial cells**: are cells that exhibit cuboidal shape, limited mobility, apical-basal polarity, and strong E-cadherin-mediated cell-cell adhesions. These cells undergo collective migration during biological processes such as morphogenesis and wound healing.

**Focal adhesion**: is a multi-protein complex that mechanically links the cell's cytoskeleton to the extracellular matrix through integrins, enabling the cell to sense forces, generate traction, and regulate migration.

**Integrin**: is a transmembrane protein that constitutes part of a focal adhesion complex.

**Mechanical stress**: is a physical quantity that describes the magnitude and direction of the force per unit area causing a system deformation.

**Membrane surface tension**: is a measure of the cohesiveness of the membrane, which is influenced by the dilational viscoelasticity of the lipid bilayer and the actin cortex, as well as by the local curvature arising from the bending of the lipid bilayer.

**Mesenchymal-like cancer cells**: are tumour cells that exhibit elongated shapes, increased migratory ability, front-rear cell polarity, and weak N-cadherin-mediated cell-cell adhesion.

**Piezo1 molecule:** is a large mechanosensitive homotrimeric membrane protein with a triskelion-like structure comprising extracellular blades, a cap, intracellular anchors, beams, and a pore.

**Strain**: is the deformation of a system in response to mechanical stress.

**Triskelion:** is an ancient symbol with three interlocking spirals, or other patterns in triplicate that share a common center.

**Ventral stress fibers:** are bundles of actin filaments and non-muscle myosin II filaments. These

> fibers are anchored at both ends to focal adhesions, and in this way they connect neighbouring focal adhesions across the ventral cytoskeletal surface. Focal adhesions serve as mechanical junctions that connect the contractile stress fibers into an integrated, load-bearing network.
>
> **Viscoelasticity**: A time-dependent response of the cell membrane under fluctuations, which includes energy storage and dissipation during its structural changes. Mechanism of energy storage and dissipation is closely connected with the stress and strain relaxation phenomena and can be described in the form of proper constitutive model.

1. Introduction

Deeper comparative analysis of the mechanosensing of epithelial and mesenchymal, cancer cells (**Glossary of terms**) offers the possibility of improving current therapies against cancer [1-4]. Cell mechanosensing involves mechanical interaction between mechanosensitive Piezo1 channels and FAs in mesenchymal-like cancer cells whereas, in epithelial cells, cell-cell adherens junctions are also involved (**Glossary of terms**). This aligns with the observation that epithelial cells create strong E-cadherin-mediated cell-cell adhesion contacts, while cancer cells form weaker N-cadherin-mediated adhesion contacts or exist as free-moving cells (**Glossary of terms**) [5]. Piezo1 is a large homotrimeric membrane protein (~2547 amino acids per monomer) with a triskelion-like structure **(Glossary of terms)** comprising extracellular blades, a cap, intracellular anchors, beams, and a pore [6-10]. It measures ~24 nm in diameter and 9 nm in depth, giving a projected area of ~450 nm², and can migrate laterally within the membrane [11]. Expression of Piezo1 is increased in the majority of cancers, with mesenchymal-like cells, when compared to healthy epithelial cells, resulting in an increased sensitivity of cancer to mechanical stress [12-14].

Mechanically-induced activation of Piezo1 molecules is responsible for influx of calcium. Local influx of calcium plays a crucial role in both the assembly and disassembly of FAs, as demonstrated by the dependence of these processes on calpain [15-20]. Epithelial cells and cancer cells display markedly different mechanisms for regulating intracellular calcium levels. Epithelial cells typically maintain intracellular calcium levels within a tightly-regulated physiological range. While calcium oscillations can also occur in normal cells under specific conditions, persistent low-frequency oscillations on the order of ~0.01 Hz have been reported in certain cancer cell lines, such as breast and prostate cancer cells, and are often associated with altered mechanosensitive and signaling states. [21,22]. The consequence of these oscillations is shorter FA lifetimes in cancer cells, which enables them to migrate faster under the same environmental conditions. Conversely, cancer cells exhibit an inability to sense matrix rigidity, resulting in considerably dynamical FAs [23]. As a result, traction force no longer scales with matrix rigidity because mechanical signals between FAs and the cytoskeleton are dissipated. The higher level of intracellular calcium and its oscillations indicate the increased activity of Piezo1 molecules in cancer cells. The activity of these molecules depends on: actomyosin contractions, cell tractions, viscoelasticity, stiffness and surface tension of the cell membrane (**Glossary of terms**). Viscoelasticity of the membrane includes constitutive behavior of the cytoskeleton and lipid bilayer, as well as their mechanical coupling under active and passive fluctuations of the membrane. In both cell types, the ventral region of the cytoskeleton is stiffer than the dorsal due to stronger actomyosin contractions. Mesenchymal-like cancer cells are more contractile than epithelial cells [24]. In cancer cells, stiffness is inhomogeneously distributed along the cytoskeleton, with the stiffest areas near long-lived FAs where stress fibers couple to cell tractions [25]. In cancer cells, ventral stress fibers are long and aligned, resulting in the formation of inhomogeneous, anisotropic structures (**Glossary of terms)** [26]. Conversely, the fibers present in epithelial cells are shorter, more cross-linked and contribute to the formation of an isotropic, more homogeneous cytoskeleton [27,28]. These physical parameters influence the tilting and footprint depth of Piezo1 molecules, which have a feedback impact on their conformational changes. Viscoelasticity and surface characteristics of the membrane influence the distribution of Piezo1 molecules.

While Piezo1 molecules are homogeneously distributed along the membrane in cancer cells, these molecules tend to group near FAs in migrating epithelial cells [29,30]. However, reduced cell contractions leads to a redistribution of Piezo1 far from FAs. The activity of Piezo1 molecules near FAs can be higher or lower than that of Piezo1 far from FAs in migrating epithelial cells [30]. However, the overall activity of Piezo1 molecules is higher in cancer cells. Although the distribution and activity characteristics of Piezo1 molecules in epithelial and cancer cells have been experimentally validated, explanations for these intricate multi-scale phenomena are lacking.

The main aim of this theoretical study is to emphasize the impact of: (i) the formation of membrane curvature adjacent to FAs on the inhomogeneous distribution of membrane surface tension which is pronounced in epithelial cells; (ii) the surface tension gradient on the distribution of Piezo1 molecules; (iii) cellular tractions, actomyosin contractility, and the distribution of mechanical stress induced by membrane fluctuations on the activity of Piezo1 molecules, which is enhanced in cancer cells; and (iv) the footprint depth of Piezo1 molecules on the interactions mediated by the membrane between Piezo1 molecules grouped near FAs in epithelial cells.

## 2. Main physical characteristics of epithelial and mesenchymal-like cancer cells responsible for the distribution of Piezo1 molecules

It is well known that Piezo1 molecules preferentially localize to inwardly curved regions of the membrane [31,32]. Such membrane curvatures can be generated around FAs depending on the distribution of ventral stress fibers, size and stability of FAs, as well as the stiffness and viscoelasticity of the substrate matrix. A thorough comprehension of the differences in the Piezo1 molecule distribution along the membrane, as well as the differences in their activity in epithelial and mesenchymal-like cancer cells, requires an appreciation of the origins of this behaviour. These include interplay between: (i) the sizes and rearrangement of contractile units within stress fibers, (ii) the rearrangement of stress fibers within the ventral region of the cytoskeleton, (iii) the contractility of the cells, (iv) the maturity/strength and stability of the FAs, (v) the traction exerted by the cells, (vi) the strength of the cell-cell adhesion contacts, and (vii) the stiffness of the substrate matrices. A comparative analysis of these physical parameters across both cell types yields a deeper understanding of what is an intricate phenomenon.

## 2.1 Arrangement of ventral stress fibers in epithelial and cancer cells

Ventral stress fibers consist of contractile units (CUs) linked sequentially in series. Each CU functions as a matrix rigidity-sensing protein complex. The sizes of these complexes vary from 2 to 3 µm. These units gather initially at the cell's periphery upon its first contact with a substrate matrix, before their organisation into stress fibers [23]. The CUs consist of myosin IIA, actin filaments, tropomyosin 2.1 (Tpm 2.1), α-actinin 4, along with various other cytoskeletal proteins [2]. The CUs in epithelial cells occur in greater number and larger size compared to those in cancer cells on the same substrate matrix. Epithelial MCF10A cells have the ability to generate CUs, while MDA-MB-231 cells lack the Tpm 2.1 cytoskeletal protein which is crucial for proper rigidity sensing [2].

The contractile actomyosin units in ventral stress fibers are mechanically coupled to FAs, depending on the stiffness of the substrate matrices (**Glossary of terms**). Ventral stress fibers connect and stabilise FAs. These fibers show heterogeneity; some extend from lateral adhesions to the leading edge, though these are usually thinner and less contractile than classic ventral fibers [26]. The fibers pull on the FA through adaptor proteins (talin, vinculin, paxillin, etc.). The FA then transmits that pulling force to integrins and thence to the substrate matrix. On rigid substrates, CUs facilitate the formation of mature FAs, while on softer substrates; the contractions exhibited by these units are very short-lived, resulting in the rapid disassembly of adhesions [23].

Healthy epithelial cells exhibit robust local contractile activity on the order of ~20–40 CUs per 100 µm² per 10 min on both soft and stiff substrates during early spreading [2,23]. In contrast, transformed cancer cells of multiple types produced <2–5 CUs per 100 µm² per 10 min on both soft and stiff substrates [23].

The arrangement of individual ventral stress fibers within the cytoskeleton plays a crucial role in the determination of: (i) the stiffness of the ventral region of the cytoskeleton and (ii) the cell contractility, which in turn influences cell tractions, as well as the sizes and strength of FAs [33-35]. Although epithelial stress fibers are composed of larger CUs compared to those found in cancer cells on an identical substrate matrix, the contractility exhibited by cancer cells is significantly greater than that of epithelial cells even on a softer substrate [24]. The rearrangement of stress fibers not only affects cell contractility, but it also impacts the viscoelastic properties of the cell membrane. Viscoelasticity of the membrane accompanied by stiffness of the cytoskeleton, cell contractility and tractions influence the distribution, conformation and activity of Piezo1 channels. The origin of the intensive contractility of cancer cells is related to the rearrangement of ventral stress fibers within the cytoskeleton [24,36].

While ventral stress fibers in cancer cells are longer, they are inhomogeneously distributed along the cytoskeleton by forming anisotropic domains. In contrast, the fibers in epithelial cells are shorter and form isotropic well-crosslinked structures [27,28]. The alignment of stress fibers in cancer cells is not random — it is driven by both intracellular signaling and extracellular mechanical feedback. However, stronger cell-cell adhesions together with FAs in epithelial cells reduce the mobility of stress fibers and their alignment. While the distribution of contractility is inhomogeneous across the cytoskeletons of cancer cells, it exhibits a more uniform distribution in epithelial cells [25].

Besides cell contractility, the arrangement of these stress fibers affects the viscoelasticity and stiffness of the ventral region of the cytoskeleton. Ventral stress fibers are attached at both ends to FAs, facilitating the connection of adjacent FAs across the ventral surface of the cell [35]. Consequently, in the ventral cytoskeletal network, FAs operate as mechanical junctions that link the contractile stress fiber bundles into a singular, load-bearing network. The latter region in mesenchymal-like cancer cells exhibits inhomogeneously distributed stiffness when compared to the ventral region of the cytoskeleton found in epithelial cells. The average stiffness of the cytoskeleton depends on: (i) the stiffness of ventral and dorsal stress fibers and intermediate filament network, (ii) the rearrangement of microtubules, and (iii) the interactions between them. It is known that the stiffness of the epithelial cytoskeleton is distributed more uniformly than that of cancer cells due to the presence of a circumferential actomyosin belt at cell–cell junctions and the keratin intermediate filament network. Ventral stress fibers in cancer cells are much stiffer and more contractile than dorsal stress fibers [37]. The keratin intermediate filament network, found in the epithelial cytoskeleton, is stiffer than the vimentin filament network presented in the mesenchymal-like cancer cytoskeleton [38]. Lorenz et al. [38] highlighted that keratin and vimentin filaments exhibit contrasting rheological behaviour under stretching: while keratin filaments extend by maintaining their stiffness, vimentin filaments become softer, undergo entropical changes under stretching, rather than entalpical changes under physiological conditions  This observation can be attributed to the distinct mechanisms of energy dissipation: the viscous sliding of subunits in keratin filaments and the non-equilibrium unfolding of α helices in vimentin filaments [38].

## 2.2 The size and stability of focal adhesions in cancer and epithelial cells

Focal adhesions (FAs) stabilize cell migration through four key characteristics: their number, size, distribution, and lifetime. These characteristics are regulated by three primary factors: (i) actomyosin contractility, (ii) intracellular calcium concentration, and (iii) substrate stiffness and viscoelasticity.

The relationship between substrate mechanics and contractility directly influences FA size. On stiffer matrices, increased contractility of actomyosin promotes protein recruitment to FAs, which causes their enlargement. In contrast, on softer matrices, reduced contractility leads to higher protein dissociation rates, ultimately causing FA disintegration [30,39]. Calcium concentration also affects FA stability, with the impact depending both on calcium levels and on the rate of concentration change [17,18,20].

FA characteristics are cell type-specific meaning that different cell types exhibit distinct FA organization patterns based on their migration behavior and stiffness of substrate matrix. For example, invasive ovarian adenocarcinoma cells (SKOV-3) grown on fibronectin matrices with Young's modulus of 25 kPa develop long-lived FAs of 3 μm in diameter with an aspect ratio of 2 [40]. Epithelial cells use both E-cadherin-mediated cell-cell contacts and FAs to support migration, while cancer cells rely primarily on FAs for stabilization. This fundamental difference leads cancer cells to form larger and more dynamic FAs compared to epithelial cells, which in turn affects migration speed [41,42].

Ozawa and Tsuruta (2011) systematically examined FA properties in epithelial cells, including NHEK keratinocytes, HaCaT cells, and 804G rat bladder carcinoma cells [41]. While mesenchymal-like cancer and epithelial cells had similar numbers of long-lived FAs, their spatial organization and size differed significantly. Cancer cells had FAs distributed throughout both the cell periphery and center, whereas epithelial cells had their FAs positioned exclusively at lateral edges to coordinate with cell-cell adhesion contacts. Additionally, epithelial cells formed consistently smaller long-lived FAs than cancer cells.

Short-lived FAs (nascent adhesions) behave differently. They are much smaller than long-lived FAs and are located near the leading edge regardless of cell type. However, cancer cells undergoing phenotypic changes show more complex patterns. Nurmagambetova et al. examined FA sizes in MCF-7A, A549, and HaCaT cells before and after the epithelial-to-mesenchymal transition [43]. While keratinocytes (HaCaT) maintained smaller FAs compared to their mesenchymal state, epithelial-like cancer cells (MCF-7A and A549) surprisingly developed larger FAs than their mesenchymal counterparts.

Despite structural differences, FA dynamics across cell types remains remarkably consistent between cell types. Ozawa and Tsuruta found no significant differences in FA dynamics between mesenchymal and epithelial cells [41]. Long-lived FAs in both cell types persist for tens of minutes, while nascent adhesions last only tens of seconds. This stability contrasts sharply with the rapid turnover of individual FA proteins, which exchange within seconds. Consequently, during a single FA's lifetime, its constituent proteins undergo multiple cycles of dissociation and re-association [44]. As a specific example, the lifetime of mature FAs in mesenchymal-like U2OS osteosarcoma cells grown on fibronectin-coated polyacrylamide substrates is approximately 50 minutes [33].

## 2.3 The perturbed membrane curvature near long-lived FAs

A long-lived FA is a multi-protein complex consisting of integrins, talin, vinculin, and actin linkers. Integrins are single-pass heterodimeric transmembrane receptors that consist of a large extracellular ligand-binding domain and short cytoplasmic tail domains. These trans-membrane proteins connect the extracellular matrix (ECM) with the cytoskeleton via proteins such as: talin, vinculin, paxillin, kindlin, zyxin, and others. Trans-membrane proteins, as inclusions, cause local bending of the lipid bilayer [45]. Son et al. (2017) discussed the impact of conformational changes of $\beta_1$ integrin, caused by cytoskeleton interactions with ECM ligands, on the lipid rearrangement and feedback impact of this rearrangement on FAs themselves [46]. Integrins form clusters in PIP2/PIP3-rich nanodomains of the lipid bilayer and sometimes associate with membrane microdomains. When integrins switch between inactive (bent) and active (extended) conformations, caused by interactions with ECM ligands, their helices can tilt or separate, creating local lipid packing mismatch. Local enrichment of cone-shaped lipids (like PIP2) at integrin clusters favors negative curvature of the bilayer (bending toward the cytoplasm). The conformational changes of integrin have the potential to regulate the thickness of the lipid bilayer [47]. Zhang et al. (2023) observed membrane curvature caused by the presence of a cluster of integrins connected to ligands of soft substrates, i.e., so called "curved adhesion" [48].

The tilted integrin and the inward bending of the bilayer alter the orientation of talin and vinculin, which are connected to the cytoskeleton, result in a "pushing" effect on adjacent actin filaments as is shown in

**Figure 1**:

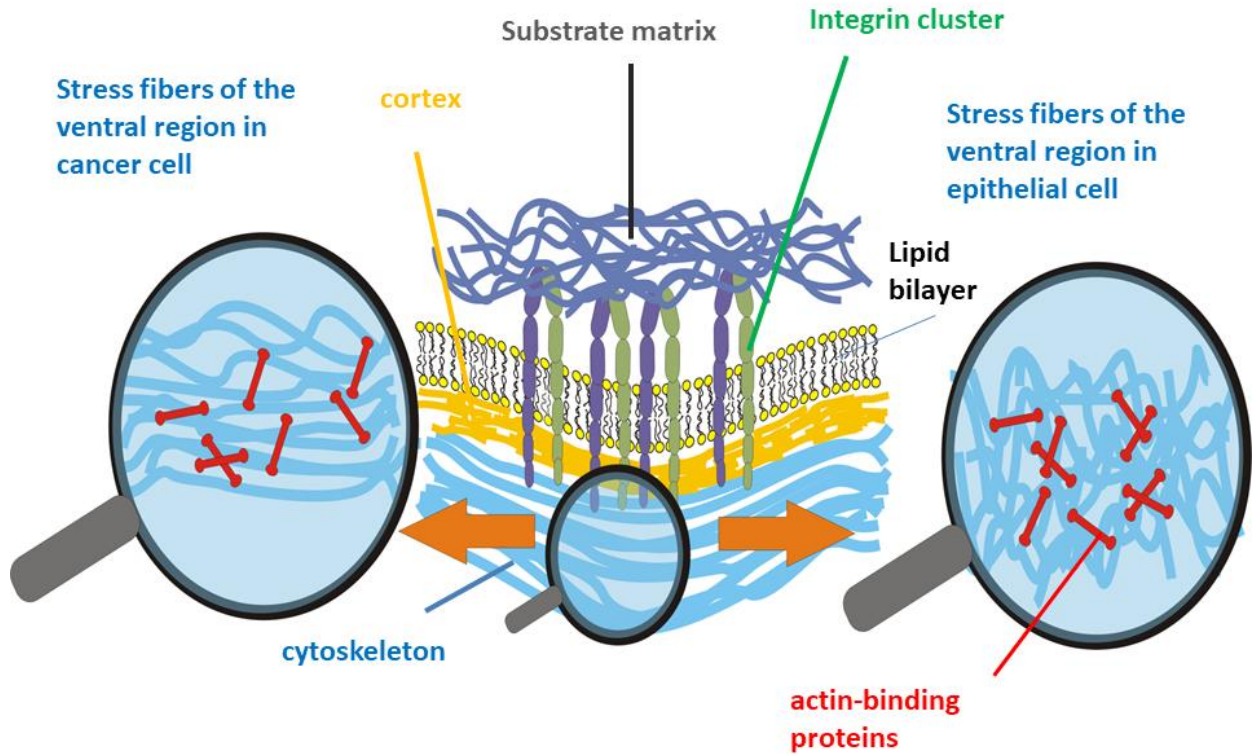

**Figure 1**. The tilting and conformational alterations of integrin clusters, driven by cell tractions and interactions between integrins and lipids, lead to bending of the lipid bilayer. This bending can consequently produce a localized inward curvature of the membrane, which is influenced by the stiffness of the cytoskeleton's ventral region. This effect is particularly evident in epithelial cells. In cancer cells, the ventral region adjacent to FAs exhibits significantly greater stiffness due to the anisotropic arrangement of stress fibers. Conversely, in epithelial cells, the stress fibers tend to form more isotropic structures.

This influences local rearrangement of ventral stress fibers [34]. In accordance with the fact that the actin cytoskeleton is pre-stressed by the myosin II contractility, even small changes in attachment geometry can impose local curvature on ventral stress fibers [35]. This recruits and orients actin filaments underneath the adhesion, shaping cytoskeletal curvature around the adhesion rim.

Consequently, the presence of matured long-lived FAs and pulling between integrin and ligands of ECM perturb the local membrane curvature and cytoskeletal stress distribution. The perturbed membrane curvature depends on: (i) size of the FA, (ii) matrix structural homogeneity, stiffness, and viscoelasticity, and (iii) the stiffness of the ventral stress fiber network. The ventral region of mesenchymal cells is stiffest near the FA due to mechanical coupling between cell tractions and actomyosin contractility. However, the ventral region of epithelial cells, which consists of shorter isotropically distributed stress fibers, is homogeneous [49]. The inhomogeneous organization of the ventral region in cancer cells generates higher resistive forces and, under the same FA pulling force, suppresses the formation of local membrane curvature, as will be discussed based on the formulated force balance of the membrane in **Section 6.3**. So even if FAs in both cell types transmit similar forces, the homogeneous ventral region of epithelial cells amplifies the membrane deformation. Although the size of the FA and the stiffness of the ECM continue to influence the local membrane curvature, when these factors remain constant, the inhomogeneity of the ventral cytoskeleton becomes the primary determinant of variations in curvature. Rearrangement of stress fibers and the remodelling of FAs affect the rearrangement of the actin cortex [50] and viscoelasticity of the ventral region of the cytoskeleton and lipid bilayer, which in turn feeds back on the distribution and activity of Piezo1 molecules.

### 3. Curvature-induced change in the membrane surface tension near focal adhesion

The cell membrane consists of two layers: the lipid bilayer and actin cortex, which are linked to the cytoskeleton (**Figure 1**). The surface tension, as a measure of the membrane cohesiveness, generally tends to reduce the membrane surface area (**Glossary of terms**). However, the relationship between the lipid bilayer and cortical surface tension depends on the context. The actin-cortex is inhomogeneous [51], pointing to an inhomogeneous distribution of the cortex surface tension. The structural composition of the cortex, and the inhomogeneous distribution of its subcellular components, exhibit considerable variation depending on the type of cell and its physiological condition [51]. Thus, cortical tension can increase the effective tension of the membrane, although the membrane tension can be higher or lower than the cortical tension depending on the specific cellular forces.

The cortex surface tension represents a measure of the cortex cohesiveness. The cortex tension of: (i) HeLa cells is $\sim 0.2 \frac{mN}{m}$ [50], (ii) F9 WT cells is a few $\frac{mN}{m}$ [52], (iii) Dictyostelium cells is $\sim 1.5 \frac{mN}{m}$ [53], (iv) E15.5 LifeAct-EGFP embryo epidermis is $\sim 0.3 \frac{mN}{m}$ [54]. The membrane surface tension includes: a dilational, viscoelastic in-plane contribution and a curvature-dependent out-of-plane contribution, which change over minutes. It can be expressed as:

$$\gamma_m(r, t_L) = \gamma_m^{in-plane} + \gamma_m^{out-of-plane} \tag{1}$$

where $r$ is the radial coordinate near the FA, $t_L$ is the timescale of minutes, $\gamma_m^{in-plane}$ is the in-plane dilational contribution and $\gamma_m^{out-of-plane}$ the curvature-dependent out-of-plane contribution to the membrane surface tension $\gamma_m$. The in-plane contribution to the membrane surface tension includes the contribution of cortex $\gamma_c^{in-plane}$ and the contribution of bilayer $\gamma_{BL}^{in-plane}$, i.e., $\gamma_m^{in-plane} = \gamma_c^{in-plane} + \gamma_{BL}^{in-plane}$. Both contributions can be expressed as: $\gamma_i^{in-plane} = K_i \frac{\Delta A}{A}$ (where $i \equiv c, BL$ and $K_i$ is the corresponding area expansion modulus). The modulus $K_{BL}$ depends on the amount of cholesterol and is equal to $K_{BL} \sim 100 \frac{mN}{m}$, while $\frac{\Delta A}{A}$ is $\sim 2-4\%$ [55]. The modulus $K_c$ is lower than the modulus $K_{BL}$ due to the fact that the actin cortex is more elastic than the bilayer. The change of in-plane contribution $\Delta \gamma_m^{in-plane}$ can be expressed in the form of a suitable constitutive model of dilational viscoelasticity $\Delta \gamma_m^{in-plane} - \frac{\Delta A}{A}$. For exact formulation of this constitutive model, additional experiments are needed. The out-of-plane, curvature-dependent contribution $\gamma_m^{out-of-plane}$ can be expressed as: $\gamma_m^{out-of-plane} = \alpha_c \frac{1}{2} |\vec{\nabla} h|^2$ (where $\alpha_c$ is the effective stiffness of the membrane, and $h(r, t_L)$ is the membrane height, which depends on the maturity (strength) of FAs $m$. Change of the maturity of FAs will be discussed in **Section 5**. Bavi et al. demonstrated that the local curvature, with a radius $\leq 50$ nm, along with the bending direction, affects the distribution of membrane surface energy and mechanical stress, which in turn has a feedback effect on the activity and distribution of Piezo1 molecules [56]. The curvature of the membrane near the FA is concave, i.e. inward shaped. An increase in the membrane surface tension is induced by a local increase in the surface area around FAs caused by curvature formation.

It means that the membrane surface tension near the FAs is lower than the surface tension far from FAs. We postulate that this inhomogeneous distribution of membrane surface tension can induce migration of the Piezo1 channels toward the FAs. The phenomenon is known as the Marangoni effect. This effect exists in various soft matter systems, but has not been discussed in the context of lateral migration of Piezo1 molecules. The surface tension gradient in soft matter systems can be induced by changes in: (i) the temperature, (ii) the distribution of constituents, or (iii) the curvature within surfaces [57]. The surface tension gradient guides the system flow/diffusion from regions of lower surface tension to regions of higher surface tension [57]. So Piezo1 molecules migrate laterally through the membrane from regions of lower membrane surface tension to regions of higher surface tension near FAs. Yang et al. discussed the distribution of Piezo1 in live cells induced by membrane curvature [31].

**4. Rearrangement of ensemble of Piezo1 molecules near focal adhesions and the Marangoni effect**

The distribution of Piezo1 molecules exhibits greater homogeneity in mesenchymal-like cancer cells; however, in epithelial cells, Piezo1 is distributed in a more heterogeneous manner [30]. A fraction of the Piezo1 molecules accumulates near FAs, whereas the remaining molecules are uniformly distributed along the membrane in these cells. The homogeneous Piezo1 distribution in cancer cells indicates that a diffusion mechanism is dominant, effecting the movement of Piezo1 molecules from regions of higher to lower surface packing density. Piezo1 molecules perform hop diffusion between two domains of the actin cytoskeleton on a time scale of milliseconds [32]. Successive hops induce damping effects in the migration of Piezo1 molecules [32]. Long-time diffusion results from the cumulative effect of successive hops. It corresponds to sub-diffusion, which may be described by use of fractional derivatives [4,58]. Actin-mediated confinement of Piezo1 diffusion does not preclude transient interactions with lipid micro-domains, but rather reflects the dominant mechanical regulation of its mobility at the membrane.

In epithelial cells, a fraction of Piezo1 molecules is situated close to FAs, indicating that lateral diffusion is not the sole mechanism responsible for the movement of Piezo1 molecules. This observation aligns with the understanding that the diffusion of Piezo1 molecules tends to produce a uniform distribution of these molecules along the membrane. An additional driving force is the gradient of the membrane surface tension, which facilitates the movement of Piezo1 molecules from regions of lower to higher surface tension adjacent to FAs, a phenomenon known as the Marangoni effect. Furthermore, the curvature of the membrane near FAs contributes to a reduction in membrane surface tension. The contributions of both mechanisms to the distribution of Piezo1 molecules along the membrane can be quantified in the form of the dimensionless Marangoni number $Ma$. The $Ma$ assesses the relative strength of surface tension-driven (Marangoni) flow in comparison to diffusive flow and can be expressed as: $Ma = \frac{\Delta \gamma_m / L}{D_\alpha \eta_\alpha L^2}$ (where $\eta_\alpha$ is the effective modulus of the membrane which accounts for anomalous energy storage and dissipation during the membrane structural changes discussed in **Section 6.3** (**Box 2**), and $L$ is the diameter of a Piezo1 molecule). Consequently, the Marangoni number for cancer cells is $Ma \leq 1$, while for migrating epithelial cells the it is $Ma > 1$.

The mass balance equation for Piezo1 molecules can be expressed as:

$$D_t^\alpha c_p(r, t_L) = -\vec{\nabla}_s \cdot \left(D_\alpha \vec{\nabla}_s c_p + \mu_p c_p \vec{\nabla}_s \gamma_m\right) \qquad (2)$$

where $c_p(r, t_L)$ is the surface number density of Piezo1 molecules, $\mu_p$ is the mobility of Piezo1 molecules driven by the gradient of the membrane surface tension, $D_\alpha$ is the anomalous diffusion coefficient in units of $\frac{m^2}{s^\alpha}$, $D_t^\alpha(\cdot)$ is a fractional derivative, and $\alpha$ is the order of the fractional derivative which satisfies the condition $0 < \alpha < 1$, described as the damping coefficient. The fractional derivatives of a function $f(t)$ are equal to $D_t^\alpha(f(t)) = \frac{d^\alpha}{dt^\alpha} f(t)$. We used Caputo's definition of the fractional derivative [58] as follows: $D_t^\alpha(f(t)) = \frac{1}{\Gamma(1-\alpha)} \frac{d}{dt} \int_0^t \frac{f(t')}{(t-t')^\alpha} dt'$ (where $t$ is the independent variable time, and $\Gamma(1-\alpha)$ is the gamma function. If the order of the fractional derivative is $\alpha = 0$ then $D_t^0(f(t)) = f(t)$. When $\alpha = 1$, the gamma function is $\Gamma(1-\alpha) \to \infty$. For this case, the fractional derivative is not defined. However, it can be shown that when $\alpha \to 1$, then $D_t^\alpha(f(t)) \to \frac{df(t)}{dt}$. The first term on the right-hand side of eq. 2 represents the diffusive flux, while the second represents the Marangoni flux. The short-term diffusion coefficient of Piezo1 molecules in erythrocyte membrane measured over the first 5 s is equal to $0.0075 \frac{\mu m^2}{s}$ [32]. Activated Piezo1 molecules conduct calcium, which has a feedback impact on the maturity/strength of the FAs themselves [30].

## 5. Long-lived focal adhesion maturation

Maturation i.e., the strength of long-lived FAs depends on: (i) the intracellular concentration of calcium, (ii) actomyosin contractility, and (iii) stiffness of a substrate matrix. Note that –

- A reduction in cell contractility enhances the disassembly of FAs, which in turn modifies the localization of Piezo1 molecules within epithelial cells, resulting in their dispersion away from FAs [30].

- An increase in matrix stiffness stimulates the maturation of FAs by: (i) increasing the strength of the bond between integrin and matrix ligands, (ii) enhancing integrin clustering and (iii) reinforcing actomyosin contractility [23]. Structural inhomogeneity of a substrate matrix causes inhomogeneous maturation of FAs.
- Epithelial cells and cancer cells exhibit significantly different mechanisms for the regulation of intracellular calcium levels. Metastatic human prostate and breast cancer cell lines, such as PC-3M and MDA-MB-231, exhibit spontaneous intracellular calcium oscillations with a frequency of approximately 0.01 Hz [21,22]. While Piezo1 activity is known to modulate calcium influx in cancer cells, its direct role in generating these oscillations was not addressed in these reports. Yao et al. [30] show that the mechanosensitive, calcium-permeable channel Piezo1 contributes to calcium entry in cells and links Piezo1's membrane recruitment with intracellular Ca²⁺ signals. In contrast to cancer cells, epithelial cells are capable of sustaining their intracellular calcium level continuously, in minutes to hours, within a consistent physiological range, so long as the cell remains viable and healthy. The migration of Piezo1 molecules is driven by: (i) the gradient of Piezo1 concentration and (ii) the gradient of membrane surface tension. Increased intracellular calcium levels in cancer cells activate calpain, a calcium-dependent protease, which can cleave talin, vinculin, and other components of FAs, resulting in their disassembly. Conversely, a moderate concentration of calcium in epithelial cells promotes the growth of FAs. As a result, a higher concentration of calcium in cancer cells, increases the specific rate of FA disassembly ($k_{diss}$.) relative to the rate of assembly ($k_{ass}$), while a moderate concentration of calcium enhances $k_{ass}$ relative to $k_{diss}$. Oscillations of intracellular calcium concentration over minutes in cancer cells additionally regulate the ratio between the assembly/disassembly of FAs.

The maturity/strength of FAs changes over minutes and can be expressed as:

$$\frac{dm}{dt_L} = k_{ass}(C_{Ca}, \vec{\Gamma}_{con}, k_F)(1-m) - k_{diss}(C_{Ca}, \vec{\Gamma}_{con}, k_F)m \quad (3)$$

where $k_{ass}$ is the specific rate of FA association and $k_{dis}$ is the dissociation rate, $C_{Ca}$ is the intracellular concentration of calcium, $k_F$ is the spring constant between integrin cluster and matrix ligands which will be discussed in **Section 6.2**, and $\vec{\Gamma}_{con}$ is the surface density of actomyosin contractions, which will be discussed in **Section 6.1**. The maturity $m$ of an FA satisfies the condition $0 < m \leq 1$, such that a matured FA is characterised as $m = 1$. While the association/dissociation rates are almost constant in epithelial cells, these rates fluctuate in cancer cells by decreasing the lifetime of FAs. Change in the calcium concentration is a multi-time process, as discussed in **Box 1**:

**Box 1**. Multi-time fluctuations of the intracellular concentration of calcium

> The calcium accumulation caused by opening a fraction of Piezo1 molecules located at $r$ within minutes can be expressed as:
>
> $$\frac{dC_{Ca}(r,t_L)}{dt_L} = k_{in}y_p - k_{out}C_{Ca} \quad (4)$$
>
> where $t_L$ is the time scale of minutes, $y_p$ is the fraction of opening Piezo1 molecules within a group of molecules located at $r$, $k_{in}$ is the specific rate of calcium inflow, $k_{out}$ is the specific rate of calcium outflow, and $C_{Ca}$, is the intracellular concentration of calcium.

The long-term change of the fraction of opening Piezo1 molecules $y_p(r, t_L)$, as a measure of their activity, can be expressed in the form of a Langevin-type equation as:

$$\frac{dy_p}{dt_L} = k_{open}(\vec{F}_{tot})(1 - y_p) - k_{close} y_p + \xi_{int} \tag{5}$$

where $\vec{F}_{tot}$ is the total periodic force surface density that acts on Piezo1 molecules, including the contributions from: (i) actomyosin contractions, (ii) cell tractions, and (iii) the distribution of mechanical stress within the membrane generated under membrane fluctuations, while $\xi_{int}$ is the stochastic force caused by hydrophobic interactions between Piezo1 molecules and lipids, $k_{open}$ is the channels' specific rate of opening $k_{open} = k_0 e^{-\frac{\Delta U_{eff}}{k_B T}}$, $\Delta U_{eff}$ is the effective potential equal to: $\Delta U_{eff} = \Delta U_0 - \Delta A \vec{F}_{tot} \cdot \Delta \vec{x}_{eff} - \Delta U_{int}$, $\Delta A$ is the membrane surface increment, $k_B$ is the Boltzmann constant, $T$ is temperature, $\Delta \vec{x}_{eff}$ is the in-plane cumulative displacement of a group of Piezo1 molecules, $\Delta U_0$ is the intrinsic double-well potential barrier between mostly-open and mostly-closed states of the group of Piezo1 molecules, and $k_{close}$ is the channels' specific rate of closing $k_{close} = k_0 e^{-\frac{\Delta U_0}{k_B T}}$, $\Delta U_{int}$ is the interaction potential expressed as:

$$\Delta U_{int} = \Delta U_{int}^0(H) + \sum_{i,j} u(r_{ij}) q_i q_j \tag{6}$$

where $\Delta U_{int}^0(H, I)$ is the baseline interaction energy of a single Piezo1 channel, modulated by the local mean curvature $H \approx \nabla^2 h$ of the membrane, $u(r_{ij})$ is the membrane-mediated coupling between two neighboring Piezo1 molecules, while $q_i$ and $q_j$ are the footprint depth of the i-th and j-th Piezo1 molecules. The baseline interaction energy can be expressed as: $\Delta U_{int}^0(H, I) = \lambda(I) H$ (where $\lambda(I)$ is a parameter that depends on the shape irregularity of the membrane curvature expressed by the irregularity index $I = \langle |\vec{\nabla} H| \rangle$ and determines how the curvature affects Piezo1 activation). If $\lambda < 0$, inward curvature raises the effective potential $\Delta U_{eff}$, which reduces the activation of Piezo1 molecules. However, if $\lambda > 0$, inward curvature lowers the effective potential $\Delta U_{eff}$, thereby enhancing the activation of Piezo1 molecules. The irregularity index depends on the homogeneity and viscoelasticity of the substrate matrices. Peussa et al. reported that Piezo1 molecules are less active during migration of epithelial Madin–Darby Canine Kidney (MDCK) cells on inhomogeneous viscoelastic PAA substrates [59]. Such substrate inhomogeneity leads to an inhomogeneous traction force distribution within individual FAs and irregular tilting of integrin clusters, resulting in irregularly shaped membrane curvatures along FAs. The combination of these irregular local curvatures and interactions between Piezo1 molecules within clusters near FAs may underlie the reduced activity of Piezo1.

The footprint depth $q$, is a physical observable that mediates membrane-mediated interactions between Piezo1 molecules, and will be discussed in **Section 7**. The average steady state condition obtained for $\frac{d\langle y_p \rangle}{dt_L} = 0$ corresponds to the fixed point equal to $\langle y_p \rangle^F = \frac{1}{1 + e^{-\frac{\Delta A \vec{F}_{tot} \cdot \Delta \vec{x}_{eff} + \Delta U_{int}}{k_B T}}}$.

The fixed point depends on the model parameters $\vec{F}_{tot}$ and $\Delta U_{int}$ and satisfies the condition that $\langle y_p \rangle^F < 1$. It is in accordance with fact that the opening of some Piezo1 molecules restricts the opening of other molecules in their neighborhoods under the same loading condition, which is above the stress threshold as previously shown for the bacterial MscL channels [60]. Interestingly, from 11 channels located within $1.6\ \mu m^2$ only 5 channels were opened at the same time [60].

Consequently, the force density $\vec{F}_{tot}$ and interaction potential $\Delta U_{int}$ can: decrease or increase the potential barrier $\Delta U_0$, thereby enhancing or decreasing the activity of Piezo1 molecules depending on the curvature size, shape, and irregularity index [56,59]. Curved regions of the lipid bilayer stimulate local accumulation of cholesterol, which can additionally reduce the activation of Piezo1 molecules near FAs [8]. The interactions between Piezo1 molecules may be overlooked when considering the case of uniformly distributed molecules along the membrane, as is the case in cancer cells. The average distance between two Piezo1 molecules in the erythrocyte membrane is 540 ± 37 nm [32]. However, the membrane-mediated interactions between Piezo1 molecules should be taken into consideration when the distance between two surrounding Piezo1 molecules corresponds to a few hundred nm or less. This condition can be satisfied for the group of Piezo1 molecules located near FAs in epithelial cells.

Thus, the interactions between Piezo1 molecules in this case can induce collective stochastic resonance depending on the characteristics of the curvature. Stochastic resonance is a physical phenomenon where the presence of a certain optimal amount of noise in a nonlinear system facilitates the crossing of an energy barrier. In many systems, weak periodic signals are too small to cross an activation threshold (e.g., energy barrier, gating threshold) [61]. Stochastic resonance in voltage-dependent ion channels caused by two-state conductance fluctuations was discussed by Goychuk and Hänggi [62]. Schmid et al. discussed collective stochastic resonance of ion channels caused by altered fluctuations of conductance. For stochastic resonance to occur through interactions among Piezo1 molecules near FAs in epithelial cells, the mean hopping time between the predominantly open and closed states of the channel cluster must match half of the effective period of the periodic driving force $\vec{F}_{tot}$ [63]. It could be expected that this condition will be satisfied in epithelial cells near FAs.

The interactions among Piezo1 molecules grouped near FA is shown in **Figure 2**:

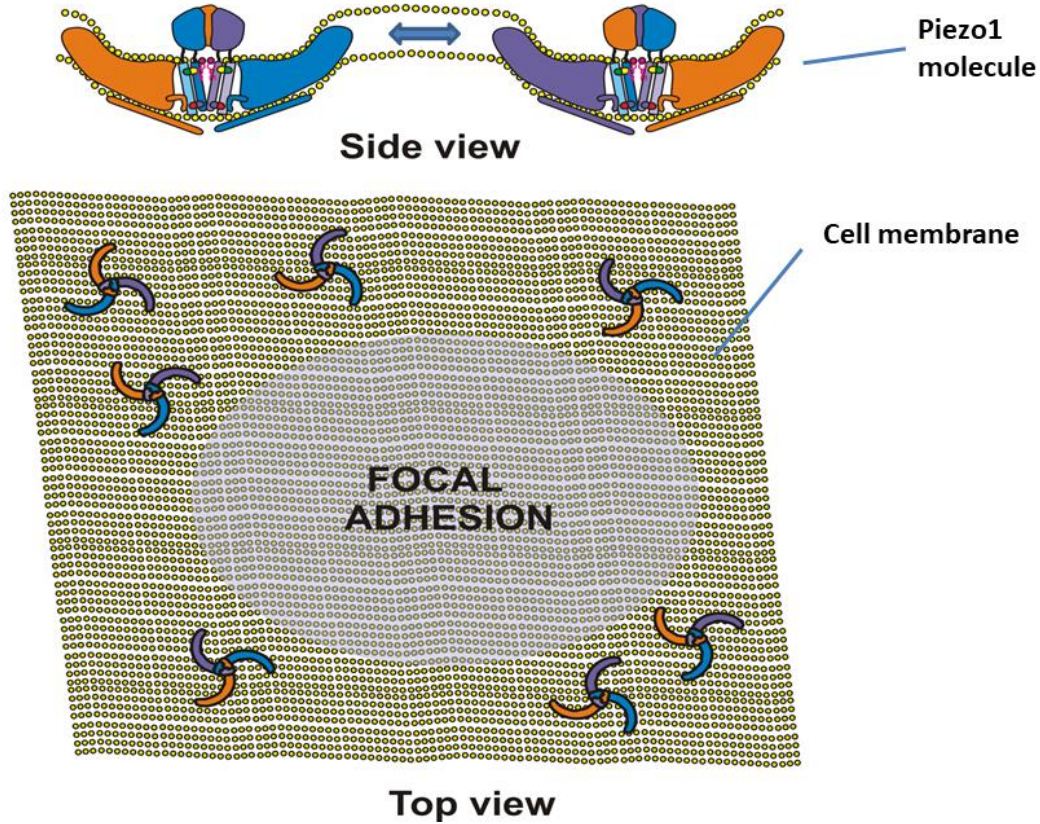

**Figure 2**. A group of Piezo1 molecules near an FA in an epithelial cell. While the average distance between Piezo1 molecules is higher in cancer cells and in epithelial cells far from FAs, this distance decreases for groups of Piezo1 molecules near FAs. In this case, it is necessary to take into consideration the interaction potential between Piezo1 molecules, which additionally decreases the height of the potential barrier separating the mostly-open from the mostly-closed states.

In further consideration, it is necessary to estimate which contribution to the driving force plays the dominant role in the activation of Piezo1 molecules.

## 6. Force responsible for the distribution and activation of Piezo1 molecules

To estimate which contribution to the force surface density $\vec{F}_{tot}$ plays the dominant role in the activation of Piezo1 molecules, we need to take into account the following set of already established facts:

- The activity of Piezo1 molecules influences the intracellular calcium concentration and its temporal variations. Evidence suggests that Piezo1 molecules exhibit increased activity in cancer cells. This observation aligns with the finding that cancer cells possess a higher intracellular

- calcium concentration compared to epithelial cells. In cancer cells, the concentration fluctuates, whereas in epithelial cells, it remains relatively stable [21,22].
- Piezo1 molecules are distributed homogeneously along the membranes of cancer cells. However, a fraction of Piezo1 molecules is grouped near FAs, while the other fraction is distributed along the membrane in epithelial cells [30]. The activity of Piezo1 molecules located near FAs in epithelial cells differs from that of Piezo1 molecules positioned at a distance from FAs, due to the influence of curvature on the conformational alterations of Piezo1 molecules, and the contributions of the traction force and the interactions between near-FA Piezo1 molecules themselves, on their activity [56,59]. The average size of an FA in migrating epithelial MDCK cells depends on the stiffness of substrate matrices and are: ∼2 $\mu m$ length and ∼1 $\mu m$ width [64]. The radius of curvature of the membrane caused by the presence of an FA should be larger than the longer size of FA.
- The contractility of cancer cells is greater than that of epithelial cells. It is in accordance with the fact that ventral stress fibers align into an anisotropic structure. Consequently, the contractile force itself is enough to activate Piezo1 molecules in cancer cells, but is insufficient to activate them in epithelial cells.
- Active and passive fluctuations of the membrane result in the generation of mechanical stress depending on the membrane's structural homogeneity and viscoelasticity. Consequently, the viscoelastic force, as the main resistive force, can be expressed as: $\vec{F}_{vis} = l_m \nabla \cdot \tilde{\sigma}_m$ (where $\tilde{\sigma}_m$ is the mechanical stress caused by the membrane active and passive fluctuations (**Glossary of terms**) and $l_m$ is the thickness of the membrane). This force relies not on the magnitude of mechanical stress that has been generated within the ventral stress fiber network and its stiffness, but rather on how mechanical stress is distributed, and it is equal to the divergence of the mechanical stress. Consequently, a pronounced inhomogeneous distribution of mechanical stress in cancer cells results in a higher value of the viscoelastic force, in comparison with epithelial cells.

The total force density $\vec{F}_{tot}$ that influences the activity of Piezo1 molecules located at $r$ is equal to:

$$\vec{F}_{tot} = \vec{\Gamma}_{con} + \sum_i \vec{F}_{t\,i} \delta(r - r_i) - \vec{F}_{vis} \quad (7)$$

where $r = r(x,y)$ is the in-plane coordinate, $\vec{\Gamma}_{con}$ is the surface density of actomyosin contractions, $\vec{F}_{t\,i}$ is the traction force exerted on the i-th FA, and $\vec{F}_{vis}$ is the viscoelastic force. The contractile force of epithelial and cancer cells satisfies the condition that $\vec{\Gamma}_{cont}^{m} > \vec{\Gamma}_{cont}^{e}$. The influence of the traction force on the activation of Piezo1 molecules within cancer cells is markedly less than that observed in epithelial cells, where Piezo1 molecules are not grouped in proximity to FAs [30]. The viscoelastic force of cancer cells is higher than that of epithelial cells $\vec{F}_{vis}^{m} > \vec{F}_{vis}^{e}$. It is a well-established fact that mechanical stress is uniformly distributed throughout the cytoskeleton of epithelial cells, whereas in the membranes of cancer cells, the distribution is inhomogeneous. High viscoelastic force generated in cancer cells is the main resistive factor that protects curvature formation along FAs. While the surface density of actomyosin contractions and the traction force are the driving forces for the membrane active fluctuations, the viscoelastic force is the resistive force.

### 6.1 Periodic contractile force

The surface density of actomyosin contractions depends primarily on the orientation of stress fibers. This force density changes over minutes and can be formulated as a coloured noise:

$$t_t \frac{d\vec{\Gamma}_{cont}}{dt_L} = -\vec{\Gamma}_{cont} + \sqrt{2D_c}\vec{\Gamma}_T \tag{8}$$

where $t_t$ is the correlation time of contractions, $D_c = t_t \langle \vec{\Gamma}_{cont}^2 \rangle$ is the variance of fluctuations and $\vec{\Gamma}_T$ is white noise caused by thermal fluctuations.

### 6.2 Traction force

Given that the traction force significantly influences the activation of Piezo1 molecules in epithelial cells, it would be worthwhile to compare the tractions exhibited by epithelial cells and cancer cells. As the size of FAs expands (or contracts), the traction forces exerted on the underlying substrate correspondingly increase (or decrease) [35]. While cancer cells are more contractile than epithelial cells, the transmission of contractile energy from the cytoskeleton of epithelial cells to FAs is more efficient than in the case of cancer cells. As a result, not all the internal contractile energy is transmitted to the substrate — a lot is "dissipated" internally in cancer cells. Traction force microscopy often finds that cancer cells exert comparable or sometimes even higher traction forces than epithelial cells [24,65]. The variability and polarisation of traction forces is higher in cancer cells and can be enhanced during cell migration on a structurally inhomogeneous substrate. Despite the fact that Piezo1 molecules are not grouped near FAs in cancer cells, the traction force generated by superposition of various FAs still influences their activity.

More matured FAs are capable of inducing stronger tractions. Stronger tractions can lead to partial disintegration of a matrix such as collagen I gel [66]. Consequently, the stiffness of a substrate matrix $k_F$ is inversely related to the maturity $m$ of an FA as $k_F \sim m^{-1}$. The phenomenon can be articulated through an appropriate constitutive model of the viscoelasticity of the extracellular matrix that connects the traction force $\vec{F}_t$ to the displacement field of the substrate matrix $\vec{u}_s$. The Zener constitutive model, suitable for viscoelastic solids, has been used for describing the viscoelasticity of various collagen matrices [67]. The traction force changes over minutes due to coupling with actomyosin contractions, and can be formulated as [67]:

$$\tau_F \frac{d\vec{F}_t(r,t_L)}{dt_L} + \vec{F}_t = k_F(m)\vec{u}_s + \eta_F(m)\frac{d\vec{u}_s(r,t_L)}{dt_L} \tag{9}$$

where $\tau_F$ is the relaxation time of the traction force under constant displacement, $k_F(m)$ is the spring constant that quantifies the strength of integrin-ligand bonds which correlates with matrix stiffness, and $\eta_F(m) = \eta_m \Delta A_{FA}$, while $\eta_m$ is the matrix viscosity, $\vec{u}_s$ is the displacement of a substrate matrix, and $\Delta A_{FA}$ is the surface area of matrix covered by the FA. The first term on the right-hand side of eq. 9 is related to energy storage, while the second term is related to energy dissipation within the matrix caused by cell tractions.

The traction force, accompanied by actomyosin contractions, induces membrane fluctuations, which generate mechanical stress. The stress distribution along the membrane generates the viscoelastic force.

### 6.3 Viscoelastic force

In cancer cells, the viscoelastic force is markedly greater owing to the inhomogeneous organization of ventral stress fibers, which results in an inhomogeneous distribution of mechanical stress along the membrane. This force, therefore, hinders the formation of membrane curvatures at FAs in cancer cells. Conversely, in epithelial cells, mechanical stress is distributed more uniformly, which is due to a more organized ventral cytoskeleton in these cells. Cell tractions, accompanied by actomyosin contractions, induce fluctuations of the cell membrane which can be described by a displacement field $\vec{u}_m(r, t_L) = \vec{u}_m(\vec{u}_c(u_{cx}, u_{cy}), h)$ (where $\vec{u}_c(x, y, t_L)$ is the in-plane displacement vector and $h = h(x, y, t_L)$ is the out-of-plane displacement). Fluctuations of the membrane generate strain $\tilde{\varepsilon}_m = \frac{1}{2}\left(\vec{\nabla} \vec{u}_m + \vec{\nabla} \vec{u}_m^T\right)$. Altered structural changes lead to energy storage and dissipation. This strain generates mechanical stress $\tilde{\sigma}_m$. The distribution of stress along the membrane is responsible for generating the viscoelastic force:

$$\vec{F}_{vis} = l_m \nabla \cdot \tilde{\sigma}_m \tag{10}$$

The stress-strain constitutive model of the membrane describes the mechanism of energy dissipation in terms of its viscoelasticity. Experimental techniques like AFM or micropipette aspiration usually probe from the dorsal region of the membrane because the ventral side is physically inaccessible under physiological condition. The dorsal region of epithelial cells is stiffer, more isotropic, and more homogeneous than that of cancer cells [4,68] The ventral region of the cytoskeleton near FAs is stiffer than the dorsal region in many cell types, due to more intense actomyosin contractions. The difference between the stiffness of ventral and dorsal regions depends on the stiffness of the substrate matrices. The ventral region of cancer cells is characterized by an inhomogeneous distribution of stiffness and mechanical stress, as well as an anisotropic alignment of stress fibers [25]. It is reasonable to propose that the rheological behaviour of both the ventral and dorsal regions satisfies the same constitutive model. We will then explore the constitutive behaviour of the dorsal region, based on microrheological experiments presented in the form of storage and loss moduli vs. frequency for a range of cell types cited in the literature [68-70]. The storage modulus represents a measure of energy storage, while the loss modulus quantifies the energy dissipation caused by the membrane structural changes under fluctuations. The range of frequencies corresponds to a timescale from minutes to seconds. A constitutive model of the dorsal region of the membrane will be discussed in **Box 2**:

**Box 2**. Constitutive model of cell membrane

> The viscoelasticity of the cell membrane depends on: (i) the viscoelasticity of the cytoskeleton, (ii) the viscoelasticity of the lipid bilayer and (iii) the way in which the cytoskeleton is coupled to the bilayer. The viscoelasticity of cytoskeleton depends on: (i) the flexibility of constitutive elements such as: stress fibers, microtubules, and intermediate filaments, (ii) the actin binding proteins, (iii) actomyosin contractility, and (iv) cell-cell and cell-matrix adhesions [4]. The viscoelasticity of the bilayer is primarily affected by its fluidity, caused by movement of lipids, while the bilayer bending has been treated as an elastic deformation [71].
>
> While the rearrangement of the bilayer leads to anomalous energy storage/dissipation caused by sub-diffusion of lipids, rearrangement of the cytoskeleton leads to both regular and anomalous energy storage/dissipation under membrane fluctuations. Regular energy storage occurs within cytoskeletal filaments and inter-filament bonds, while energy is dissipated by the breaking of some bonds during conformational changes of the filaments. Anomalous energy storage/dissipation within the cytoskeleton is caused by the establishment and breaking of short-lived bonds. This anomalous energy storage and dissipation, associated with the damped structural alterations of the cytoskeleton and bilayer, has been characterized using fractional derivatives.
>
> The lamellar structure of cell membrane leads to parallel mechanical coupling between the cytoskeleton and bilayer [72,73]. It means that the membrane fluctuation-induced strain is the same within the cytoskeleton and the bilayer, while the membrane in-plane stress represents the sum of stresses generated within the cytoskeleton and bilayer. The constitutive model of the membrane formulated by Pajic-Lijakovic et al. represents a type of fractional Kelvin-Voigt model expressed as [4]:
>
> $$\widetilde{\boldsymbol{\sigma}}_m(r, t_L) = G_{SC}\widetilde{\boldsymbol{\varepsilon}}_m + \eta_\alpha D_t^\alpha \widetilde{\boldsymbol{\varepsilon}}_m + \eta_C \dot{\widetilde{\boldsymbol{\varepsilon}}}_m \qquad (11)$$
>
> where $\widetilde{\boldsymbol{\sigma}}_m(r,t_L)$ and $\widetilde{\boldsymbol{\varepsilon}}_m(r,t_L)$ are respectively the in-plane stress and strain within the membrane, $D_t^\alpha(\cdot)$ corresponds to the fractional derivative, $\alpha$ is the order of fractional derivative which satisfies the condition $0 < \alpha < 1$ [73], $G_{SC}$ is the elastic modulus of the cytoskeleton, $\eta_C$ is the viscosity of the cytoskeleton, $\eta_\alpha$ is the effective modulus of the membrane as a whole, and $\dot{\widetilde{\boldsymbol{\varepsilon}}}_m = \frac{d\widetilde{\boldsymbol{\varepsilon}}_m}{dt_L}$.
>
> The first term on the right-hand side of eq. 10 describes regular energy storage, while the third term quantifies regular energy dissipation within the cytoskeleton. The second term quantifies the anomalous nature of energy storage and dissipation within both the cortex and bilayer.

The membrane fluctuations can be formulated based on the in-plane and out-of-plane force balances. The in-plane force balance can be expressed as:

$$\frac{1}{\Delta A}\xi_u \frac{d\vec{u}_c}{dt_L} = \vec{\nabla}\gamma_m^{in-plane} + \vec{F}_{tot}^{\ tan} \qquad (12)$$

where $\vec{F}_{tot}^{\ tan}$ is the tangential component of the total force expressed by eq. 6, equal to $\vec{F}_{tot}^{\ tan} = (\vec{t}\cdot\vec{F}_{tot})\cdot\vec{t}$, $\vec{t}$ is the unit vector in tangential direction, and $\xi_u$ is the in-plane drag coefficient. The tangential components of the traction force and the surface density of actomyosin contractions (components of $\vec{F}_{tot}^{\ tan}$), are the driving forces, while the surface tension force (the first term on the right-hand side (RHS) of eq. 11) and the tangential component of the viscoelastic force (components of $\vec{F}_{tot}{}^{tan}$) are the resistive forces for membrane in-plane fluctuations.

The out-of-plane displacement can be expressed as:

$$\frac{1}{\Delta A}\xi_h\frac{dh}{dt_L} = \sum_{j}^{N_{FA}} F_{FAi}^n(q_{FA})\delta(r_i - r) - \kappa\nabla^4 h + \vec{\nabla}\cdot\left(\gamma_m^{out-of-plane}\vec{\nabla}h\right) + \sum_{i=1}^{N} F_{pi}(q)\,\delta(r_i - r) + F_{tot}{}^n$$

(13)

where $\kappa$ is the bending modulus of the lipid bilayer, $\xi_h$ is the drag coefficient for the out-of-plane movement of the membrane, $F_{FAi}^n(q_{FA}) = -\frac{\partial E_m}{\partial q_{FA}}$ is the integrin-induced normal force density, $q_{FA}\sim\Delta h$ is the footprint depth of an FA, $N_{FA}$ is the number of FAs, $F_{tot}^n$ is the normal component of the surface force density of cell tractions presented in scalar form as: $F_{tot}^n = \vec{n}\cdot\vec{F}_{tot}$, while $\vec{n}$ is the unit vector in the direction perpendicular to the membrane, and $F_{pi} = -\frac{\partial E_m}{\partial q}$ is the normal force caused by the presence of Piezo1 (i.e., an inclusion), $E_m$ is the free energy of the membrane, which will be described in **Section 7**, $q$ is the footprint depth of Piezo1, and $N$ is the number of Piezo1 molecules located at $r$. The normal components of the traction force and surface density of actomyosin contractions (components of $F_{tot}{}^n$), the integrin-induced normal force density, and the Piezo1-induced normal force density are driving forces, while the bending force density (the second term on the RHS of eq. 13), the surface tension force (the third term on the RHS of eq. 13) and the normal, i.e., out-of-plane component of the viscoelastic force (components of $F_{tot}{}^n$) are the resistive forces for membrane out-of-plane fluctuations. When driving forces are higher than resistive forces, curvatures can be formed around FAs. The integrin-induced normal force density constitutes the primary driving force in contrast; the normal component of the viscoelastic force is the leading resistive force.

Membrane fluctuations affect the footprint depth of Piezo1 molecules, which in turn provides feedback on membrane-mediated interactions between neighboring Piezo1 molecules.

## 7. Membrane fluctuations-induced change in the footprint depth of a Piezo1

The footprint depth of Piezo1 molecules depends on: (i) the free energy of the membrane, and (ii) the driving force $\vec{F}_{tot}$. The footprint depth is lower in the stiffer ventral regions than in the dorsal and lateral regions of the membrane. The free energy $E_m$ of the membrane is made up of several contributions, of which the most important are:

- The Helfrich types of bending energy of the bilayer near Piezo1 per unit area $E_B = \frac{\kappa}{2}\left(\frac{2}{R(q)} - C_0^p\right)^2$, where $\kappa$ is the bending modulus, which accounts for the bilayer and cortex contributions, $C_0^p$ is the spontaneous curvature, $R(q)$ is the curvature of cup-like deformation of the membrane near Piezo1 equal to $R(q) = \frac{r_f^2 + q^2}{2q}$, $r_f$ is the radius of the base of the spherical cup, while $q$ is the depth of the cup. The bending modulus increases with an increase in the concentration of cholesterol within the bilayer [74]. An increase in the bending modulus resists an increase in the footprint depth $q$ [75,76].

- The surface energy $E_A$ per unit area is equal to $E_A = \gamma_m$ (where $\gamma_m$ is the membrane surface tension, which accounts for in-plane and out-of-plane contributions). The membrane surface tension does work in reducing the cortex surface area caused by the presence of a Piezo1 molecule (as an inclusion), which reduces the increase in footprint depth $q$.

- The strain energy $E_{str}$ per unit area is equal to: $E_{str} = l_m \frac{1}{2} \widetilde{\boldsymbol{\sigma}}_{cm} : \widetilde{\boldsymbol{\varepsilon}}_{cm}$ (where $l_m$ is the membrane thickness). An increase in the strain energy caused by the membrane fluctuations resists further increase in the footprint depth.

For describing the dynamics of Piezo1, it is necessary to formulate the balance of thermodynamic affinities as a Langevin-type equation for change of the Piezo1 footprint depth, which occurs over minutes:

$$\frac{1}{\Delta A(q)} \gamma \frac{dq(t_L)}{dt_L} = -\frac{\partial E_m}{\partial q} + F_{tot}^n \tag{14}$$

where $q$ is the footprint depth, $\gamma$ is the friction coefficient, $E_m$ is the free energy of the membrane per unit area caused by structural changes during fluctuations, $E_m(q) = E_B + E_A + E_{str}$, $\Delta A(q)$ is the area difference relative to the flat base disk of radius $r_f$ equal to: $\Delta A(q) = A_{cup}(q) - \pi r_f^2$, while $A_{cup}(q)$ is the cup area equal to $A_{cup}(q) = 2\pi q R(q)$. The membrane free energy $E_m$ increases with increasing $q$. In general, the resistance effects of the membrane, as measured by the membrane free energy, are reduced in cancer cells. Conversely, the driving force is increased, which results in an increased Piezo1 footprint depth. A higher footprint depth ensures a higher open probability of Piezo1 molecules.

**8. Interrelationship between physical parameters governing the distribution and activity of Piezo1 channels**

Cellular mechanosensing—the detection of mechanical cues such as stiffness, stress, or topography and their conversion into biochemical signals—regulates processes like morphogenesis, tissue homeostasis, and disease. In epithelial and mesenchymal-like cancer cells, mechanosensing is strongly linked to Piezo1 channel distribution and activity. This distribution is governed by (i) Piezo1 surface concentration and (ii) the gradient of the membrane surface tension, as shown in **Figure 3**:

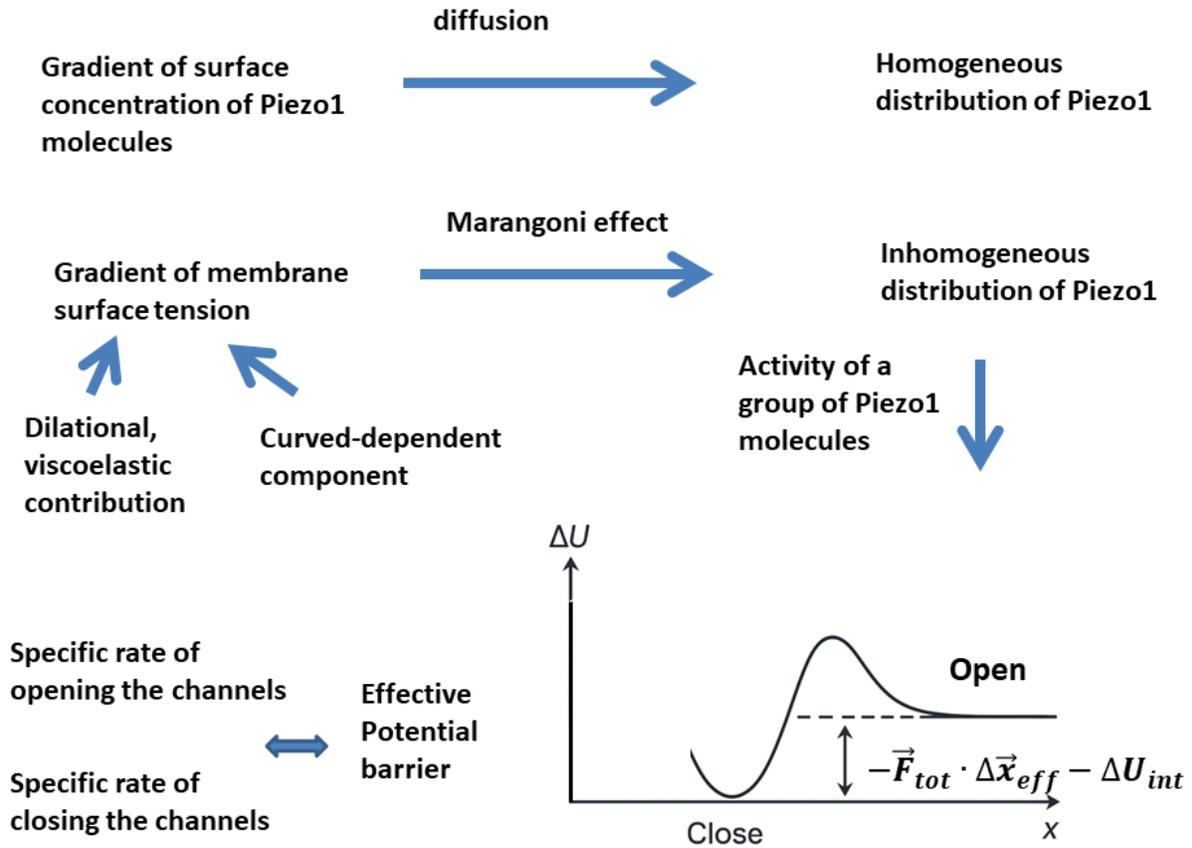

**Figure 3**. Schematic representation of the relationship between physical parameters that influence the distribution and activity of a group of Piezo1 molecules in migrating epithelial and cancer cells.

Initially, it is essential to examine the distinctions between cancer cells and epithelial cells in relation to the physical parameters introduced, followed by a focus on the primary physical factors that affect the distribution and activity of Piezo1 molecules in both types of cells. The differences between cancer and epithelial cells in the context of the physical parameters involved in this theoretical analysis are shown in **Table 1**:

**Table 1**. The differences between migrating cancer and epithelial cells in the context of the physical parameters introduced

| Physical parameters | Epithelial cells | Cancer mesenchymal-like cells |
|---|---|---|
| Properties of FA | Stabile, smaller | Unstable, larger |
| Calcium concentration near FA | Within the physiological range | Oscillates |
| Number of Piezo1 molecules | Smaller | Higher |
| Distribution of Piezo1 molecules | Inhomogeneous | Homogeneous |
| Piezo 1 molecules grouped near FAs | Yes | No |
| Curvature formation around FAs | Larger | Smaller |
| Curvature-induced gradient of membrane surface tension | Higher | Lower |
| Viscoelastic force | Lower | Higher |
| Traction force | Inhomogeneous | Homogeneous |

| Contractile force | Lower | Higher |
| Average activity of Piezo1 molecules | Lower | Higher |

The stability of FAs, quantified through their maturity $m$ (Eq. 3), depends on the intracellular calcium concentration (Eq. 4) and on the fraction of open Piezo1 channels (Eq. 5). Curvature formation around FAs results from a balance between the integrin-induced normal force and the opposing viscoelastic force (Eq. 13). The integrin-induced normal force (a driving force) depends on the size and stability of FAs. In contrast, the viscoelastic force provides mechanical resistance and depends on the spatial distribution of membrane stress, which is particularly pronounced in cancer cells. The resulting curvature—its size and geometry—modulates the distribution and activity of Piezo1 molecules. Larger inward curvature promotes the accumulation of Piezo1 molecules in the vicinity of FAs, a phenomenon especially evident in epithelial cells. This feedback mechanism is mediated by curvature-induced changes in membrane surface tension (Eq. 1). The change in a membrane's surface tension depends on its viscoelasticity and curvature (**Figure 1**).

The distribution of mechanical stress is determined by the structural organization of ventral stress fibers, whereas membrane surface tension depends on the rearrangement of the actin cortex and lipid bilayer. In cancer cells, the inhomogeneous organization of ventral stress fibers produces a correspondingly inhomogeneous distribution of mechanical stress, leading to an increase in resistive viscoelastic forces. Stress fibers do not contract uniformly across their entire length; rather, they function as localized, contractile, and viscoelastic bundles capable of exerting significant tugging forces on both the extracellular matrix and the cytoskeleton [84]. This leads to an inhomogeneous distribution of stress within the ventral cytoskeleton, an effect that is pronounced in cancer cells. By contrast, membrane surface tension is primarily influenced by curvature formation around FAs, which locally increases the membrane surface area and on that basis the surface tension. In cancer cells, this curvature formation is largely suppressed by the viscoelastic force, leading to a more homogeneous distribution of cortex–bilayer membrane surface tension. Consequently, mechanical stress in the ventral cytoskeleton is more inhomogeneous, whereas membrane surface tension is more uniform in cancer cells compared with epithelial cells, consistent with the suppression of curvature-induced local membrane stretching. In epithelial cells, curvature formation along FAs causes the generation of surface tension gradients that drive Piezo1 migration toward higher surface tension regions near FAs via the Marangoni effect. Because of non-uniform stress distribution the curvatures observed in cancer cells are significantly reduced as a result of the substantial resistance exerted by viscoelastic forces [84]. The reduced curvature is the main cause of the relatively uniform distribution of membrane surface tension in cancer cells. Consequently, Piezo1 molecules are redistributed primarily by diffusion, rather than by Marangoni-driven flows.

Piezo1 molecules exhibit a high propensity for clustering within curved membrane regions across diverse cell types [8,30,32,36]. These curved environments modulate the conformational states of Piezo1, creating a feedback loop that governs their mechanical sensitivity [56]. Specifically, the coupling of membrane 'footprints'—the local curvature deformations surrounding the protein—becomes more pronounced in dense clusters, subsequently influencing footprint geometry, lipid rearrangement, and channel activity. While some studies suggest that dense packing may lead to the self-inactivation of mechanosensitive channels [8,60], others contend that these channels function as independent mechanotransducers even at high cluster densities [85,86]. To distinguish between active and silent populations within these clusters, researchers have supplemented patch-clamp electrophysiology with structural and optical techniques, including Small-Angle Neutron Scattering (SANS), neutron reflection, Atomic Force Microscopy (AFM), Total Internal Reflection Fluorescence (TIRF), and Stochastic Optical Reconstruction Microscopy (STORM) [8,60,85]. Notably, some of these modalities can confound results by externally influencing channel activity; for instance, the mechanical force exerted by an AFM probe or the steric hindrance of fluorescent tags used in TIRF and STORM can alter the native conformational states and mechanical sensitivity of the channels.

The activity of Piezo1 molecules was discussed depending on the magnitude of the force $\vec{F}_{tot}$ and the interaction potential $\Delta U_{int}$ (eq. 6) that can reduce the potential barrier between the mostly-open and mostly-closed states of a group of Piezo1 molecules by increasing their specific rate of activation. The interactions among Piezo1 molecules are pronounced in epithelial cells near FAs. These lipid-mediated interactions are influenced by the footprint depth of Piezo1 $q$, presented by eq. 14. Successive, force-induced channel openings regulate intracellular calcium oscillations, which in turn modulate FA assembly and disassembly, quantified by FA maturity.

## 9. The implications of different Marangoni-driven behavior in epithelial versus cancer cells

Beyond FA recruitment, the differential Marangoni-driven behaviour of epithelial versus cancer cells represents an underappreciated layer of mechanobiological divergence. Marangoni flows arise from surface tension gradients at interfaces. In cells, such gradients can emerge from heterogeneities in membrane lipid composition, protein crowding, curvature-generating deformation, and cytoskeletal coupling. Non-transformed epithelial cells typically maintain higher cortical tension, more ordered lipid domains, and robust integrin-mediated anchorage to the extracellular matrix (ECM), creating spatially constrained surface tension landscapes as reported previously [4]. In contrast, many cancer cells exhibit reduced membrane order, altered lipid composition (including increased unsaturated lipids and cholesterol redistribution) and decreased cortical stiffness, all of which can amplify or spatially redistribute membrane surface tension gradients. Such changes contribute to enhancement of lateral membrane flows, increase in asymmetric protein redistribution, dynamic protrusion formation and mechanically biased migration. Mechanical membrane properties are central to these distinctions between the two cell types. Epithelial cells characteristically display higher effective membrane surface tension and stronger membrane–cortex coupling, partly mediated by ERM proteins and organized actin cortices, which stabilize membrane curvature and resist spontaneous interfacial instabilities [77]. Cancer cells frequently show decreased cortical stiffness and altered viscoelastic behaviour, facilitating bleb formation and curvature-driven flows that can reinforce invasive phenotypes. Membrane surface tension has been shown to regulate mechanosensitive channel activity, including that of Piezo1, which responds to bilayer surface tension and curvature contribution [78,79]. In epithelial cells, localized increases in membrane mechanical stress at FAs can spatially recruit and activate Piezo1, coupling traction forces to calcium influx and FA maturation [30]. In cancer cells lacking this recruitment, Marangoni-like membrane flows driven by local membrane surface tension gradients may become uncoupled from stabilizing calcium feedback, which can favor protrusive dynamics over adhesion reinforcement. Consequently, the loss of Piezo1 recruitment to FAs in most cancer cells is a defining characteristic of the cancerous state, influencing how cancer cells interact with their physical environment. This has significant implications for understanding cancer pathophysiology, including the loss of mechanical feedback control and the promotion of cell motility.

More specifically, integrin organization and FA density further modulate these interfacial mechanics. Normal epithelial cells display well-structured FAs enriched in integrins (e.g., β1, β3), talin, vinculin, and paxillin, forming strong mechanical linkages between ECM and actin stress fibers. These structures generate spatially confined traction stresses that shape membrane surface tension gradients locally. In contrast, many cancers cells exhibit fewer, more transient FAs and altered integrin expression profiles, contributing to reduction of mechanical constraint and enhanced migratory plasticity [80]. The absence of stable FA–Piezo1 coupling diminishes mechanochemical feedback loops that normally regulate adhesion turnover and cell spreading. Consequently, surface tension gradients may redistribute membrane proteins, including curvature-sensing BAR-domain proteins, toward leading edges, promoting lamellipodia or bleb-based motility modes. Cytoskeletal architecture also shapes Marangoni-relevant surface tension heterogeneities. Epithelial cells typically exhibit organized actomyosin networks with apico-basal polarity and junctional complexes, constraining lateral membrane flows and distributing stress isotropically and non-randomly. Cancer cells often display cytoskeletal remodelling, including reduced stress fiber organization and increased actin branching via actin-related protein 2/3 (Arp2/3) complex, alongside enhanced contractility in specific subpopulations [81]. These changes can create dynamic, spatially shifting stress across the membrane. Given that Piezo1 activity is sensitive to both bilayer surface tension and cytoskeletal force transmission [82], its abnormal localization in cancer cells can decouple membrane surface tension sensing from adhesion-mediated reinforcement, which enables persistent tension asymmetries that drive polarized membrane flows and invasive protrusions. Furthermore, membrane curvature formation represents another key element defining differences between the epithelial and cancer cells. Epithelial cells tightly regulate curvature through coordinated lipid composition, cortical tension, and curvature-stabilizing proteins, thereby limiting spontaneous blebbing and suppressing large-scale interfacial instabilities. In cancer cells, reduced membrane–cortex adhesion and altered lipid order increase susceptibility to curvature amplification. Such curvature differentials can themselves generate surface tension gradients, reinforcing Marangoni-type flows that redistribute membrane proteins and signaling complexes. The feedback between curvature, membrane surface tension, and mechanosensitive channel activation as reported for Piezo1 [83], is fundamentally rewired in cancer cells.

In summary, the divergent mechanical membrane properties, integrin–cytoskeletal connectivity, FA number and stability, and Piezo1 localization patterns existing between epithelial and mesenchymal-like cancer cells suggest that Marangoni-driven interfacial behaviors are not merely biophysical curiosities but are likely to constitute functionally significant regulators of mechanotransductive signaling, adhesion dynamics, and migratory plasticity during tumor progression and metastatic dissemination. This has significant implications for understanding cancer pathophysiology, including the loss of mechanical feedback control and the promotion of cell motility. In addition, the localization of Piezo1 is also proposed as a potential diagnostic biomarker and the protein's linker domain (residues 1418–1656) has been identified as a potential therapeutic target [30].

## 10. Conclusion

The activity and distribution of Piezo1 molecules, as well as the maturity and strength of FAs, are the key determinants of cell mechanosensing. Interestingly, migrating epithelial cells and cancer mesenchymal-like cells display markedly different behaviours in relation to these factors. In cancer cells, Piezo1 molecules are uniformly distributed, whereas in epithelial cells, their distribution is inhomogeneous. In epithelial cells, Piezo1 molecules tend to group near FAs, a phenomenon that is accentuated by actomyosin contractility, and which enhances the strength of FAs, whereas a decrease in contractility leads to a more uniform distribution of Piezo1 molecules. The main physical factor responsible for the different distribution of Piezo1 molecules in epithelial and cancer cells is the resistive out-of-plane viscoelastic force. This force arises from the inhomogeneous distribution of mechanical stress along the ventral cytoskeleton and has the potential to suppress curvature formation around FAs. In cancer cells, structural inhomogeneity of the ventral cytoskeleton leads to strongly inhomogeneous stress distributions and, consequently, to a higher viscoelastic resistive force. As a result, this force significantly reduces the formation of inward membrane curvature around FAs. In contrast, the ventral cytoskeleton of epithelial cells is more homogeneous, generating a lower viscoelastic force that is insufficient to resist curvature formation, thereby allowing inward curvature to develop around FAs. Curvature formation causes inhomogeneous distribution of membrane surface tension such that curved regions have higher surface tension in comparison with surrounding regions. Induced surface tension gradient, which is a component of the Marangoni effect, stimulates the migration of Piezo1 molecules from regions of lower surface tension, situated away from FAs, towards regions of higher surface tension in the proximity to FAs. The diffusion mechanism is another factor that affects the movement of Piezo1 molecules on the surface of epithelial cells. In cancer cells, Piezo1 molecules move laterally primarily by diffusion, migrating from regions of higher concentration to regions of lower concentration. This diffusion mechanism facilitates the formation of a uniform distribution of Piezo1 molecules.

The expression and activity of Piezo1 molecules are significantly increased in cancer cells as compared to epithelial cells. The activity of Piezo1 molecules is associated with the intracellular calcium concentration. While epithelial cells maintain stable intracellular calcium concentration, this concentration oscillates in cancer cells. The influx of calcium near FAs in epithelial cells serves to stabilize these FAs thus prolonging their lifespan. Conversely, the oscillation of intracellular calcium in cancer cells leads to the destabilization of FAs and a corresponding reduction in their lifespan. Although the differences in the behaviour of epithelial and cancer cells have been thoroughly investigated, the underlying causes of these discrepancies have remained unexplained. The primary objective of this theoretical analysis has been to provide potential explanations for these differences.

The activity of a group of Piezo1 molecules is influenced by: (i) the driving force, which has the capacity to reduce the potential barrier between the predominantly-open and predominantly-closed states of a group of Piezo1 molecules, and (ii) lipid-mediated interactions among adjacent molecules, along with conformational alterations of Piezo1 molecules induced by membrane curvature, which may either elevate or lower the potential barrier, resulting in the deactivation or activation of Piezo1 molecules. The driving force encompasses: contractile, traction forces, while the viscoelastic force is a resistive force. As a result, the increased force produced in cancer cells ensures a greater activity of Piezo1 molecules.

The activity of Piezo1 molecules located near FAs in epithelial cells can be greater or lower than that of those situated away from FAs, a phenomenon attributed to the interactions among Piezo1 molecules in proximity to FAs and curvature-induced conformation changes of Piezo1 molecules. The interaction potential can reduce or increase the potential barrier depending on the curvature shape and size. Should the average hopping time between the predominantly open and closed states of a group of Piezo1 molecules be roughly half of the effective period of the periodic driving force, collective stochastic resonance may occur, thereby enhancing the activity of Piezo1 molecules near FAs in epithelial cells.

**Conflict of interest**: The authors report no conflict of interest.

**Funding**: This work was supported in part by the Engineering and Physical Sciences Research Council, United Kingdom (grant number EP/X004597/1), by the Ministry of Science, Technological Development and Innovation of the Republic of Serbia (Contract No. 451-03-34/2026-03/ 200135) and National Health and Medical Research Council of Australia (Investigator Grant L3 2034293).